\documentclass[12pt]{article}
\usepackage{graphicx}

\renewcommand{\bar}[1]{\overline{#1}}
\newcommand{\half}{{$\frac{1}{2}$}} 
\newcommand{\VEV}[1]{\left\langle{#1}\right\rangle}
\newcommand{\bm}[1]{\mbox{\boldmath $#1$}}
\newcommand{\slsh}[1]{\mbox{$\not\! #1$}}
\begin{document}

\begin{flushright}
SLAC-PUB-9563 \\
USM-TH-133 \\
November 2002
\end{flushright}

\vspace{6pt}

\begin{center}
{\Large \bf Single Hadronic-Spin Asymmetries in Weak Interaction
Processes} \footnote{\baselineskip=13pt
Work partially supported by the Department of Energy, contract
DE--AC03--76SF00515, by the LG Yonam Foundation, and by Fondecyt
(Chile) under grant 8000017.}
\end{center}

\centerline{ \bf Stanley J. Brodsky$^a$, Dae Sung Hwang$^{b}$,
and Ivan Schmidt$^{c}$}

\vspace{8pt} {\centerline{$^a$Stanford Linear Accelerator
Center,}}

{\centerline{Stanford University, Stanford, California 94309,
USA}}

\centerline{e-mail: sjbth@slac.stanford.edu}

\vspace{8pt} {\centerline{$^{b}$ Department of Physics, Sejong
University, Seoul 143--747, Korea}}

\centerline{e-mail: dshwang@sejong.ac.kr}

\vspace{8pt} {\centerline {$^{c}$Departamento de F\'\i sica,
Universidad T\'ecnica Federico Santa Mar\'\i a,}}

{\centerline {Casilla 110-V, 
Valpara\'\i so, Chile}}

\centerline{e-mail: ischmidt@fis.utfsm.cl }

\vfill

\centerline{Submitted to Physics Letters B.}
\vfill

\newpage

\setlength{\baselineskip}{13pt}

\bigskip

\begin{abstract}
We show that measurements of single-spin asymmetries (SSAs) in
charged current weak interaction processes such as deep inelastic
neutrino scattering on a polarized target and inclusive $W$
production in polarized hadron-hadron collisions discriminate
between the two fundamental QCD mechanisms (the Sivers and Collins
effects) which have been proposed to explain such
time-reversal-odd asymmetries.  It has recently been shown that
QCD final-state interactions due to gluon exchange between the
struck quark and the proton spectators in semi-inclusive deep
inelastic lepton scattering will produce non-zero Sivers-type
single-spin asymmetries which survive in the Bjorken limit.  We
show that this QCD final-state interaction produces identical SSAs
in charged and neutral current reactions.   Furthermore, the
contribution of each quark to the SSA from this mechanism is
proportional to the contribution of that quark to the polarized
baryon's anomalous magnetic moment. In contrast, the Collins
effect contribution to SSAs depends on the transversity
distribution of quarks in the polarized target.  Since the charged
current only couples to quarks of one chirality, it cannot sense
the transversity distribution of the target, and thus it gives no
Collins-type contribution to single-spin correlations.
\end{abstract}

\bigskip

\section{Introduction}

Spin correlations provide a remarkably sensitive window to
hadronic structure and basic mechanisms in QCD.   Among the most
interesting polarization effects are single-spin azimuthal
asymmetries (SSAs) in semi-inclusive deep inelastic scattering,
representing the correlation of the spin of the proton target and
the virtual photon to hadron production plane: $\vec S_p \cdot
\vec q \times \vec p_H.$ Such asymmetries are time-reversal odd,
but they can arise in QCD through phase differences in different
spin amplitudes.

The most common explanation of the pion electroproduction
asymmetries in semi-inclusive deep inelastic scattering is that
they are related to the transversity distribution of the quarks in
the hadron $h_{1}$~\cite{Jaffe:1996zw,Boer:2001zw,Boer:2002xc}
convoluted with the transverse momentum dependent fragmentation
function $H^\perp_1$, the Collins function, which gives the
distribution for a transversely polarized quark to fragment into
an unpolarized hadron with non-zero transverse
momentum~\cite{Collins93,Barone:2001sp,Ma2002}.

Recently, an alternative physical mechanism for the azimuthal
asymmetries has been proposed \cite{BHS,Collins,Ji:2002aa}.
It was shown that the QCD final-state interactions (gluon
exchange) between the struck quark and the proton spectators in
semi-inclusive deep inelastic lepton scattering can produce
single-spin asymmetries which survive in the Bjorken limit.  In
this case, the fragmentation of the quark into hadrons is not
necessary, and one has a correlation with the production plane of
the quark jet itself $\vec S_p \cdot \vec q \times \vec p_q.$  This
final-state interaction mechanism provides a physical explanation within
QCD of single-spin asymmetries.  The required matrix element measures the
spin-orbit correlation $\vec S \cdot \vec L$ within the target hadron's
wavefunction, the same matrix element which produces the anomalous
magnetic moment of the proton, the Pauli form factor, and the
generalized parton distribution $E$ which is measured in deeply virtual
Compton scattering.  Physically, the final-state interaction phase arises
as the infrared-finite difference of QCD Coulomb phases for hadron
wavefunctions with differing orbital angular momentum.

A related analysis also predicts that the initial-state
interactions from gluon exchange between the incoming quark and
the target spectator system lead to leading-twist single-spin
asymmetries in the Drell-Yan process $H_1 H_2^\updownarrow \to
\ell^+ \ell^- X$ \cite{Collins,BHS2}. These final- and
initial-state interactions can be identified as the path-ordered
exponentials which are required by gauge invariance and which
augment the basic light-front wavefunctions of
hadrons~\cite{Collins,Ji:2002aa}.  Initial-state interactions also
lead to a $\cos 2 \phi$ planar correlation in unpolarized
Drell-Yan reactions \cite{Boer,Gamberg}.

In this paper we  extend the analysis of QCD final-state
interactions to SSAs which can be measured in weak interaction
processes. For example, consider charged current neutrino
semi-inclusive deep inelastic scattering, where a hadron (pion) is
measured in the final state.  In this case, the transversity
distribution cannot contribute to the cross section since the
produced quark from the weak interaction of the $W$ boson is
always left-handed.  This point has also been noted by
Miyama~\cite{Miyama:1999yp} and Boer~\cite{Boer:2001zw}. On the
other hand, in the final-state interaction picture the single-spin
asymmetry in charged and neutral current weak interactions will be
the same as in the electromagnetic case.  Thus these weak
interaction processes will clearly distinguish the underlying
physical mechanisms which produce target single-spin asymmetries.

Measurements of SSAs in semi-inclusive neutrino deep inelastic
scattering on a polarized target will be experimentally possible
with the advent of a muon storage ring which can provide a high
intensity well-focused  neutrino beam~\cite{Geer:2002pv}. However,
there are also other weak interaction processes in which similar
SSAs should be present and which can be  measured experimentally
at existing facilities. For example, in this paper we will also
make testable predictions for other SSAs in weak interaction
reactions such as the processes $p p^\updownarrow \to Z X,$ $p
p^\updownarrow \to W X,$ which can be measured at
RHIC~\cite{Zykunov:2001mn}, and $e^+ e^- \to Z \to \pi
\Lambda^\updownarrow X,$ where the correlation of the $\Lambda$
polarization with the production plane can be measured.

\section{SSA in Electromagnetic Interactions}

The final-state interaction effects can be identified with the
gauge link which is present in the gauge-invariant definition of
parton distributions~\cite{Collins}.  When the light-cone gauge is
chosen, a transverse gauge link is required. Thus in any gauge the
parton amplitudes need to be augmented by an additional eikonal
factor incorporating the final-state interaction and its
phase~\cite{Ji:2002aa,Belitsky:2002sm}. The net effect is that it
is possible to define transverse momentum dependent parton
distribution functions which contain the effect of the QCD
final-state interactions.  We will use this description in this
section.  It has been shown that the same final-state interactions
are responsible for the diffractive component to deep inelastic
scattering, and that they play a critical role in nuclear
shadowing phenomena~\cite{Brodsky:2002ue}.

The quark distribution in the proton is described by a
correlation matrix:
\begin{equation}
\Phi^{\alpha\beta} (x,{\bm p}_{\perp}) = \int {d\xi^-d^2{\bm
\xi}_{\perp}\over (2\pi)^3}e^{ip\cdot \xi}
\VEV{P,S|{\bar{\psi}}^{\beta}(0)\psi^{\alpha}(\xi)|P,S}\mid_{\xi^+=0}\
, \label{w1}
\end{equation}
where $x=p^+/P^+$.
We use the convention $a^{\pm}=a^0\pm a^3$,
$a\cdot b={1\over 2}(a^+b^-+a^-b^+)-{\bm a}_{\perp}\cdot {\bm b}_{\perp}$.
The correlation matrix $\Phi$ is parameterized in terms of the
transverse momentum dependent quark distribution functions \cite{MB98}:
\begin{eqnarray}
\Phi (x , {\bm p}_{\perp}) &=& {1\over 2}{\Bigg[}f_1{\slsh{n}}+
f_{1T}^{\perp} {{\epsilon}_{\mu\nu\rho\sigma}{\gamma}^{\mu}
n^{\nu}p_{\perp}^{\rho}S_{\perp}^{\sigma}\over M}
+g_{1s}{\gamma}_5{\slsh{n}}
\label{w2}\\
&&+h_{1T} i{\gamma}_5{\sigma}_{\mu\nu}n^{\mu}S_{\perp}^{\nu}
+h_{1s}^{\perp}
{i{\gamma}_5{\sigma}_{\mu\nu}n^{\mu}p_{\perp}^{\nu}\over M}
+h_1^{\perp} {{\sigma}_{\mu\nu}p_{\perp}^{\mu}n^{\nu}\over M}
{\Bigg]}\ , \nonumber
\end{eqnarray}
where the distribution functions have arguments $x$ and ${\bm p}_{\perp}$
such as $f_1(x , {\bm p}_{\perp})$, and
$n^\mu =(n^+,n^-,{\bm n}_{\perp})
=(0,2,{\bm 0}_{\perp})$.
The quantity $g_{1s}$ (and similarly $h_{1s}^{\perp}$ and
$G_{1s}$, $H_{1s}^{\perp}$ in (\ref{w4}) below)
is shorthand for
\begin{equation}
g_{1s}(x , {\bm p}_{\perp})=\lambda g_{1L}(x , {\bm p}_{\perp}^2)
+{{\bm p}_{\perp}\cdot {\bm S}_{\perp}\over M}g_{1T}(x , {\bm p}_{\perp}^2)
\ ,
\label{g1s}
\end{equation}
with $M$ the mass, $\lambda =MS^+/P^+$ the light-cone helicity, and
${\bm S}_{\perp}$ the transverse spin of the target hadron.
Integrating over ${\bm p}_{\perp}$ gives the distribution functions
$f_1(x)=\int d^2{\bm p}_{\perp}f_1(x , {\bm p}_{\perp})$,
$g_1(x)=g_{1L}(x)$, and $h_1(x)=h_{1T}(x)+h_{1T}^{\perp (1)}(x)$,
where $h_{1T}^{\perp (1)}(x)$ is the first ${\bm p}_{\perp}^2/2M^2$
moment of $h_{1T}^{\perp}(x , {\bm p}_{\perp})$ \cite{MB98}.

The quark fragmentation is described by the correlation matrix given by
\begin{equation}
\Delta^{\alpha\beta} (z , {\bm k}_{\perp}) = \sum_{X} \int
{d\xi^+d^2{\bm \xi}_{\perp}\over 2z(2\pi)^3}e^{ik\cdot \xi}
\VEV{0|\psi^{\alpha}(\xi)|X;P_h,S_h}
\VEV{X;P_h,S_h|\bar{\psi}^{\beta}(0)|0}\mid_{\xi^-=0}\ ,
\label{w3}
\end{equation}
where $z=P_h^-/k^-$ and ${\bm k}_{\perp}$ is the quark transverse momentum
with respect to the produced hadron and then $-z{\bm k}_{\perp}$ is the
produced hadron transverse momentum with respect to the fragmenting quark.
The correlation matrix $\Delta$ is parameterized in terms of the
transverse momentum dependent quark fragmentation functions \cite{MB98}:
\begin{eqnarray}
\Delta (z , {\bm k}_{\perp}) &=& {1\over 2}{\Bigg[}D_1{\slsh{\bar
n}}+ D_{1T}^{\perp} {{\epsilon}_{\mu\nu\rho\sigma}{\gamma}^{\mu}
{\bar n}^{\nu}k_{\perp}^{\rho}S_{h\perp}^{\sigma}\over M_h}
+G_{1s}{\gamma}_5{\slsh{\bar n}}
\label{w4}\\
&&+H_{1T} i{\gamma}_5{\sigma}_{\mu\nu}{\bar
n}^{\mu}S_{h\perp}^{\nu} +H_{1s}^{\perp}
{i{\gamma}_5{\sigma}_{\mu\nu}{\bar n}^{\mu}k_{\perp}^{\nu}\over
M_h} +H_1^{\perp} {{\sigma}_{\mu\nu}k_{\perp}^{\mu}{\bar
n}^{\nu}\over M_h} {\Bigg]}\ , \nonumber
\end{eqnarray}
where the fragmentation functions have arguments $z$ and $-z{\bm k}_{\perp}$
like $D_1(z , -z{\bm k}_{\perp})$, and
${\bar n}^\mu =({\bar n}^+,{\bar n}^-,{\bm {\bar n}}_{\perp})
=(2,0,{\bm 0}_{\perp})$.

The hadronic tensor of the leptoproduction by the electromagnetic
interaction in leading order in $1/Q$ is given by \cite{MB98}
\begin{eqnarray}
2M{\cal W}^{\mu\nu}(q,P,P_h)&=& \int d^2{\bm p}_{\perp} d^2{\bm
k}_{\perp} \delta^2({\bm p}_{\perp}+{\bm q}_{\perp}-{\bm
k}_{\perp})
\nonumber\\
&&\times {1\over 4}\ {\rm Tr}\Big[ \Phi (x_B,{\bm
p}_{\perp})\gamma^\mu \Delta (z_h,{\bm k}_{\perp})\gamma^\nu \Big]
\nonumber\\
&&+\ \Bigl(\ q\leftrightarrow -q\ ,\ \ \mu \leftrightarrow \nu\
\Bigr)\ ,
\label{w5}
\end{eqnarray}
where $x_B=Q^2/2P\cdot q$ and $z_h=P\cdot P_h/P\cdot q$.
The momentum ${\bm q}_{\perp}$ is the transverse momentum of the
exchanged photon in the frame where $P$ and $P_h$ do not have
transverse momenta.

The single-spin asymmetry (SSA) in semi-inclusive deep inelastic
scattering (SIDIS) $ep^{\updownarrow}\to e'\pi X$, which is given by
the correlation ${\vec S}_p\cdot {\vec q}\times {\vec p}_{\pi}$, is
obtained from (\ref{w5}).  For the electromagnetic interaction,
there are two mechanisms for this SSA: $h_1
H_1^{\perp}$ and $f_{1T}^{\perp}D_1$.  The former was first studied
by Collins~\cite{Collins93}, and the latter by
Sivers~\cite{Sivers};  it was shown recently that the Sivers effect
does not vanish in the Bjorken limit~\cite{BHS}.

We have calculated \cite{BHS} the single-spin asymmetry in
semi-inclusive electroproduction $\gamma^* p^{\updownarrow} \to H
X$ induced by final-state interactions in a model of a spin-\half
~ proton of mass $M$ with charged spin-\half ~ and spin-0
constituents of mass $m$ and $\lambda$, respectively, as in the
QCD-motivated quark-diquark model of a nucleon.  The basic
electroproduction reaction is then $\gamma^* p \to q (qq)_0$.  In
fact, the asymmetry comes from the interference of two amplitudes
which have different proton spin but couple to the same final
quark spin state, and therefore it involves the interference of
tree and one-loop diagrams with a final-state interaction. In this
simple model the azimuthal target single-spin asymmetry $A^{\sin
\phi}_{UT}$ is given by \cite{BHS}
\begin{eqnarray}
A^{\sin \phi}_{UT} &=& {C_F \alpha_s(\mu^2) } \ { \Bigl(\ \Delta\,
M+m\ \Bigr)\ r_{\perp}\over \Big[\ \Bigl( \ \Delta\, M+m\
\Bigr)^2\
+\ {\bm r}_{\perp}^2\ \Big]}\nonumber \\
&\times& \Bigg[\ {\bm r}_{\perp}^2+\Delta
(1-\Delta)(-M^2+{m^2\over\Delta} +{\lambda^2\over 1-\Delta})\
\Bigg] \nonumber\\[1ex] &\times& \ {1\over {\bm r}_{\perp}^2}\
{\rm ln}{{\bm r}_{\perp}^2 +\Delta
(1-\Delta)(-M^2+{m^2\over\Delta}+{\lambda^2\over 1-\Delta})\over
\Delta (1-\Delta)(-M^2+{m^2\over\Delta}+{\lambda^2\over
1-\Delta})}\ . \label{sa2b}
\end{eqnarray}
Here $r_\perp$ is the magnitude of the transverse momentum of the
current quark jet relative to the virtual photon direction, and
$\Delta=x_{Bj}$ is the usual Bjorken variable. To obtain
(\ref{sa2b}) from Eq. (21) of \cite{BHS}, we used the
correspondence ${|e_1 e_2|\over 4 \pi} \to C_F \alpha_s(\mu^2)$
and the fact that the sign of the charges $e_1$ and $e_2$ of the
quark and diquark are opposite since they constitute a bound
state. It has been shown that the result (\ref{sa2b}) corresponds
to the $f_{1T}^{\perp}D_1$ mechanism \cite{Collins,Ji:2002aa}. The
formula (\ref{sa2b}) is equivalent to $-\, (r_{\perp}^1\,
f_{1T}^{\perp}(\Delta , {\bm r}_{\perp}))\, /\, (M\, f_{1}(\Delta
, {\bm r}_{\perp}))$.

\begin{figure}[htbp]
\centering
\includegraphics[height=5.25in]{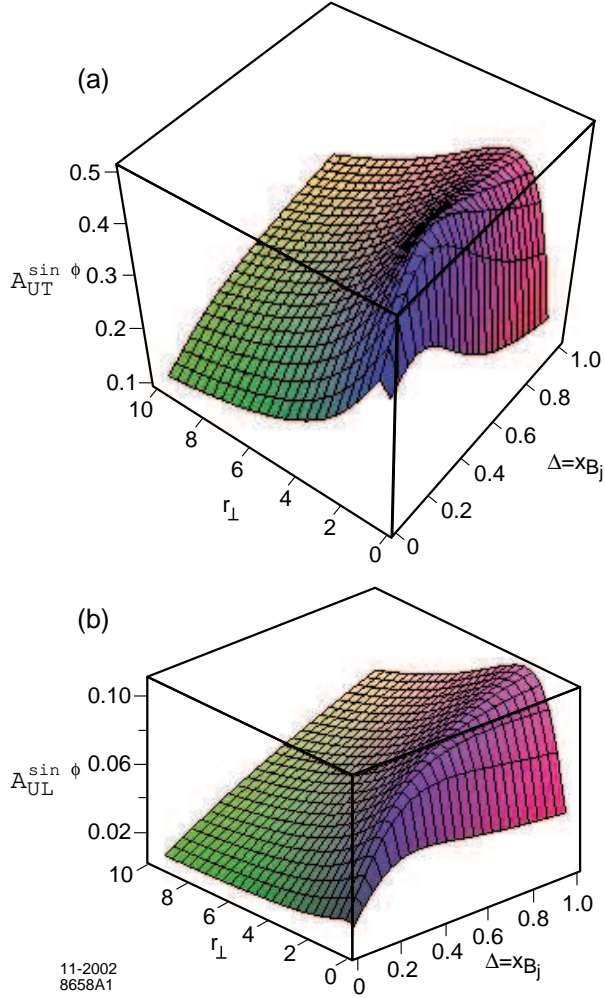}
\caption[*]{\baselineskip=12pt Model predictions for the target
single-spin asymmetry $A^{\sin \phi}_{UT}$ for charged and neutral
current deep inelastic scattering resulting from gluon exchange in
the final state.  Here $r_\perp$ is the magnitude of the
transverse momentum of the outgoing quark relative to the photon
or vector boson direction, and $\Delta = x_{bj}$ is the light-cone
momentum fraction of the struck quark. The parameters of the model
are given in the text. In (a) the target polarization is
transverse to the incident lepton direction. The asymmetry in (b)
$A^{\sin \phi}_{UL} = K A^{\sin \phi}_{UT}$ includes a kinematic
factor $K = {Q\over \nu}\sqrt{1-y}$ for the case where the target
nucleon is polarized along the incident lepton direction. For
illustration, we have taken $K= 0.26 \sqrt x,$ corresponding to
the kinematics of the HERMES experiment~\cite{Airapetian:1999tv}
with $E_{lab} = 27.6 ~{\rm GeV}$ and $y = 0.5.$} \label{fig:1}
\end{figure}

We show in Fig. \ref{fig:1}  the predictions of our model for the
asymmetry $A^{\sin \phi}_{UT}$ of the  ${\vec S}_p \cdot \vec q
\times \vec p_q$ correlation based on  Eq. ({\ref{sa2b}}).  As
representative parameters we take $\alpha_s = 0.3$, $M =  0.94$
GeV for the proton mass,  $m=0.3$ GeV for the fermion constituent
and $\lambda = 0.8$ GeV for the spin-0 spectator.  The single-spin
asymmetry $A^{\sin \phi}_{UT}$ is shown as a function of $\Delta$
and $r_\perp$ (GeV) in Fig. \ref{fig:1}(a) . The asymmetry
measured at HERMES~\cite{Airapetian:1999tv} $A_{UL}^{\sin \phi} =
K A^{\sin \phi}_{UT}$ contains a kinematic factor $K = {Q\over
\nu}\sqrt{1-y} = {\sqrt{2Mx\over E}}{\sqrt{1-y\over y}}$ because
the proton is polarized along the incident electron direction. The
resulting prediction for $A_{UL}^{\sin \phi}$ is shown in Fig.
\ref{fig:1}(b) . Note that $\vec r = \vec p_q - \vec q$ is the
momentum of the current quark jet relative to the photon momentum.
The asymmetry as a function of the pion momentum $\vec p_\pi$
requires a convolution with the quark fragmentation function.

Since the same matrix element controls the Pauli form factor,
the contribution of each quark current to the SSA is
proportional to the contribution $\kappa_{q/p}$ of that quark to the
proton target's anomalous magnetic moment
$\kappa_p = \sum_q e_q \kappa_{q/p}$~\cite{BHS}.

The SSA of the Drell-Yan processes, such as $\pi p^{\updownarrow}\
({\rm or}\ p p^{\updownarrow}) \to \gamma^* X\to \ell^+\ell^- X$,
is related to initial-state interactions.  The simplest way to get
the result is applying crossing symmetry to the SIDIS processes.
This was done in Ref. \cite{BHS2} with the result that the SSA in
the Drell-Yan process is the same as that obtained in SIDIS, with
the appropriate identification of variables, but with the opposite
sign~\cite{Collins,BHS2}.  This result corresponds to the
$f_{1T}^{\perp}f_1$ mechanism.  The SSA of Drell-Yan processes can
also arise from the $h_1 h_1^{\perp}$ mechanism.

We can also consider the SSA of $e^+e^-$ annihilation processes
such as $e^+e^-\to \gamma^* \to \pi {\Lambda}^{\updownarrow} X$.
The $\Lambda$ reveals its polarization via its decay $\Lambda \to
p \pi^-$.  The spin of the $\Lambda$ is normal to the decay plane.
Thus we can look for a SSA through the T-odd correlation
$\epsilon_{\mu \nu \rho \sigma} S^\mu_\Lambda p^\nu_\Lambda
q^\rho_{\gamma^*} p^\sigma_{\pi}$.  This is related by crossing to
SIDIS on a $\Lambda$ target.  The SSA of this process can arise
from the $H_1 H_1^{\perp}$ and $D_{1T}^{\perp}D_1$ mechanisms.

\section{SSA in Weak Interactions}

\subsection{Charged Current}

Let us consider the SSA in the charged current (CC) weak
interaction process $\nu p^{\updownarrow}\to \ell\pi X$.  For the
CC weak interaction, the trace in (\ref{w5}) becomes
\begin{equation}
{\rm Tr}\Big[ \Phi \gamma^\mu P_L \Delta \gamma^\nu
P_L \Big]
=
{\rm Tr}\Big[ \Phi P_R\gamma^\mu P_L
\Delta P_R \gamma^\nu
P_L \Big]
=
{\rm Tr}\Big[ \Phi_{\rm CC} \gamma^\mu \Delta_{\rm CC} \gamma^\nu \Big]
\ ,
\label{w6}
\end{equation}
where
$P_L=(1-\gamma_5)/2$, $P_R=(1+\gamma_5)/2$, and
\begin{equation}
\Phi_{\rm CC}\equiv P_L \Phi P_R\ ,\qquad
\Delta_{\rm CC}\equiv P_L \Delta P_R\ .
\label{w7}
\end{equation}

When we use $\Phi (x , {\bm p}_{\perp})$ and $\Delta (z , {\bm k}_{\perp})$
given in (\ref{w2}) and (\ref{w4}), (\ref{w7}) gives
\begin{eqnarray}
\Phi_{\rm CC} (x , {\bm p}_{\perp})&=& {1\over 2}\ P_L\
{\Bigg(}f_1{\slsh{n}}+ f_{1T}^{\perp}
{{\epsilon}_{\mu\nu\rho\sigma}{\gamma}^{\mu}
n^{\nu}p_{\perp}^{\rho}S_{\perp}^{\sigma}\over M}
+g_{1s}{\gamma}_5{\slsh{n}} {\Bigg)} \ ,
\label{cc1}\\[1ex]
\Delta_{\rm CC} (z , {\bm k}_{\perp}) &=& {1\over 2}\ P_L\
{\Bigg(}D_1{\slsh{\bar n}}+ D_{1T}^{\perp}
{{\epsilon}_{\mu\nu\rho\sigma}{\gamma}^{\mu} {\bar
n}^{\nu}k_{\perp}^{\rho}S_{h\perp}^{\sigma}\over M_h}
+G_{1s}{\gamma}_5{\slsh{\bar n}} {\Bigg)} \ . \nonumber
\end{eqnarray}

We see from Eq. (\ref{cc1}) that $\Phi_{\rm CC}$ does not contain
the chiral-odd distribution functions which are present in
(\ref{w2}), and $\Delta_{\rm CC}$ does not contain the chiral-odd
fragmentation functions present in (\ref{w4}). The charged current
only couples to a single quark chirality, and thus it is not
sensitive to the transversity distribution. Thus  SSAs can only
arise in  charged current weak interaction SIDIS from the Sivers
FSI mechanism $f_{1T}^{\perp}D_1$ in leading order in $1/Q$; in
contrast, both the Collins  $h_1 H_1^{\perp}$ and Sivers
$f_{1T}^{\perp}D_1$ mechanisms contribute to SSAs for the
electromagnetic and neutral current (NC) weak interactions.

We can also consider the SSAs of the processes $\pi
p^{\updownarrow}\ ({\rm or}\ p p^{\updownarrow})\to W X\to \ell
\nu X.$  Again these SSAs arise from the $f_{1T}^{\perp}f_1$
mechanism, but not from the $h_1 h_1^{\perp}$ mechanism.

\subsection{Neutral Current}

Let us now consider the SSA in the neutral current weak
interaction process $\nu p^{\updownarrow}\to \nu\pi X$.  For the
NC weak interaction, the interaction vertex of $Z$-f-f is given by
$(-ie/{\rm sin}{\theta}_{\rm W}{\rm cos}{\theta}_{\rm W})
(c_LP_L+c_RP_R)$ with the weak isospin-dependent coefficients
$c_{L,R}=I^3_{\rm W}-Q{\rm sin}^2{\theta}_{\rm W}$.  Explicit
values of $c_{L,R}$ are given by $c_L={1\over 2}-{2\over 3}{\rm
sin}^2{\theta}_{\rm W}$, $c_R=-{2\over 3}{\rm sin}^2{\theta}_{\rm
W}$ for u, c, t quarks, and $c_L=-{1\over 2}+{1\over 3}{\rm
sin}^2{\theta}_{\rm W}$, $c_R={1\over 3}{\rm sin}^2{\theta}_{\rm
W}$ for d, s, b quarks.

The trace in (\ref{w5}) becomes
\begin{eqnarray}
&&a\ {\rm Tr}\Big[ \Phi \gamma^\mu (c_LP_L+c_RP_R) \Delta \gamma^\nu
(c_LP_L+c_RP_R) \Big]
\label{nc1}\\
&=&
a\ {\rm Tr}\Big[ \Phi (c_LP_R+c_RP_L) \gamma^\mu
\Delta \gamma^\nu
(c_LP_L+c_RP_R) \Big]
\ =\
a\ {\rm Tr}\Big[ \Phi_{\rm NC} \gamma^\mu \Delta \gamma^\nu \Big]
\ ,
\nonumber
\end{eqnarray}
where $a=1/{\rm sin}^2{\theta}_{\rm W}{\rm cos}^2{\theta}_{\rm W}$
and
\begin{equation}
\Phi_{\rm NC}\equiv
(c_LP_L+c_RP_R)\Phi (c_LP_R+c_RP_L)\ .
\label{nc2}
\end{equation}
When we use $\Phi (x , {\bm p}_{\perp})$
given in (\ref{w2}), (\ref{nc2}) gives
\begin{eqnarray}
\Phi_{\rm NC} (x , {\bm p}_{\perp}) &=& {1\over 2}\ {\Bigg[}\
(c_L^2P_L+c_R^2P_R)\ {\Bigg(}f_1{\slsh{n}}+ f_{1T}^{\perp}
{{\epsilon}_{\mu\nu\rho\sigma}{\gamma}^{\mu}
n^{\nu}p_{\perp}^{\rho}S_{\perp}^{\sigma}\over M}
+g_{1s}{\gamma}_5{\slsh{n}}{\Bigg)}
\label{ncw2}\\
&&+\ c_Lc_R\ {\Bigg(}h_{1T}
i{\gamma}_5{\sigma}_{\mu\nu}n^{\mu}S_{\perp}^{\nu} +h_{1s}^{\perp}
{i{\gamma}_5{\sigma}_{\mu\nu}n^{\mu}p_{\perp}^{\nu}\over M}
+h_1^{\perp} {{\sigma}_{\mu\nu}p_{\perp}^{\mu}n^{\nu}\over M}
{\Bigg)}\ {\Bigg]}\ . \nonumber
\end{eqnarray}

For the $f_{1T}^{\perp}D_1$ mechanism, we put the former
parentheses part of (\ref{ncw2}) into (\ref{nc1}) and then we have
${\rm Tr}[\Phi^{\rm f}\gamma^\mu (c_L^2 P_L + c_R^2 P_R)\Delta
\gamma^\nu ]$, where $\Phi^{\rm f}$ is the first three terms of
$\Phi$ in (\ref{w2}).  Then, we find that the SSA is given by that
of the electromagnetic case with $f_{1T}^{\perp}D_1$ replaced by
\begin{equation}
a\ {c_L^2 + c_R^2\over 2}\
f_{1T}^{\perp}D_1\ .
\label{nc12}
\end{equation}
However, $f_1$ is also weighted by the same factor $a\, (c_L^2 +
c_R^2) / 2$, as we can see in (\ref{ncw2}). Therefore, the SSA
from the final-state interaction mechanism in the NC weak
interaction is the same as that in the electromagnetic
interaction. This can be confirmed in the simple quark-diquark
model.

For the $h_1 H_1^{\perp}$ mechanism, we put the latter parentheses part
of (\ref{ncw2}) into (\ref{nc1}) and find that the SSA is given
by that of the electromagnetic case with $(h_1 H_1^{\perp})/(f_1 D_1)$
replaced by
\begin{equation}
{2c_Lc_R \over c_L^2 + c_R^2}\ {h_1 H_1^{\perp} \over f_1 D_1}
\ .
\label{nc11}
\end{equation}
That is, the SSAs are modified by the quark weak isospin-dependent
factor $2c_Lc_R / (c_L^2 + c_R^2)$ in comparison with the
electromagnetic case. The same factor appears in the linear $\cos
\theta$ forward-backward asymmetry in the $ e^+ e^- \to Z \to q
\bar q$ reaction.

The SSA of the Drell-Yan processes at the  $Z^0$, such as $\pi
p^{\updownarrow}\ ({\rm or}\ p p^{\updownarrow})\to Z X\to
\ell^+\ell^- X$, can arise from the $h_1 h_1^{\perp}$ and
$f_{1T}^{\perp}f_1$ mechanisms. We can also consider the SSA of
the $e^+e^-$ annihilation processes such as $e^+e^-\to Z \to \pi
{\Lambda}^{\updownarrow} X$, which can arise from the $H_1
H_1^{\perp}$ and $D_{1T}^{\perp}D_1$
mechanisms~\cite{Boer:1997qn}. The SSAs of these processes have
the same situation as those of the above SIDIS case. The
initial/final-state interaction mechanisms have the same formulas
as the electromagnetic case, whereas the Collins mechanisms are
weighted by the quark weak isospin-dependent factor $2c_Lc_R /
(c_L^2 + c_R^2)$ present in (\ref{nc11}).

\section{Conclusions}

We have shown that target single-spin asymmetries in
semi-inclusive deep inelastic scattering $\nu p^{\updownarrow}\to
\ell\pi X$ from the charged current weak interaction can arise
only from the $f_{1T}^{\perp}D_1$ mechanism in leading order in
$1/Q$, whereas the SSAs in $ep^{\updownarrow}\to e'\pi X$ of
electromagnetic interaction from the $h_1 H_1^{\perp}$ and
$f_{1T}^{\perp}D_1$ mechanisms. Thus charged current weak
interaction processes clearly distinguish the underlying physical
mechanisms responsible for target single-spin asymmetries.

We have also analyzed the SSAs in semi-inclusive reactions such as
$\nu p^{\updownarrow}\to \nu \pi X$ of the neutral current weak
interaction and have found that the SSA from the Collins mechanism
is dependent on the quark weak isospin. The phase from the QCD
final-state interaction mechanism only depends on color; it is the
same as that in the electromagnetic case and does not depend on
the quark weak isospin. Furthermore, the contribution of each
quark current to the SSA from this mechanism is proportional to
the contribution of that quark to the polarized baryon's anomalous
magnetic moment.

It is also important to study the SSAs in weak interaction
reactions such as $p p^\updownarrow \to Z X,$ $p p^\updownarrow \to W
X,$ which can be measured at RHIC, and $e^+ e^- \to Z \to \pi
\Lambda^\updownarrow X$ which can be measured in $e^+ e^-$
colliders. In each case the contribution from each quark to the
SSAs from initial/final-state interactions is identical to that of
the corresponding electromagnetic process.

\end{document}